\documentclass{article}

\usepackage[english]{babel}
\usepackage[utf8x]{inputenc}
\usepackage[T1]{fontenc}
\usepackage[title]{appendix}
\usepackage{framed}
\usepackage{color}
\usepackage{algorithm}
\usepackage{algpseudocode}
\usepackage{float}
\usepackage{enumerate}
\usepackage{enumitem}
\usepackage[a4paper,top=2.5cm,bottom=2cm,left=2.75cm,right=2.75cm,marginparwidth=1.75cm]{geometry}
\usepackage{amsmath}
\usepackage{amssymb}
\usepackage{amsthm}
\usepackage{proof}
\usepackage{graphicx}
\usepackage{subcaption}
\usepackage[colorinlistoftodos]{todonotes}
\usepackage[colorlinks=true, allcolors=blue]{hyperref}
\usepackage{wrapfig,lipsum}
\usepackage{authblk}
\usepackage{multirow}
\usepackage{hhline}
\usepackage{setspace}
\usepackage{palatino}
\usepackage{mathdots}
\usepackage{multirow}
\usepackage{siunitx}
\usepackage{xcolor}
\usepackage{makecell}
\usepackage{booktabs,colortbl, array}

\newcolumntype{C}[1]{>{\centering\arraybackslash}p{#1}}

\newcommand\ket[1]{\ensuremath{|#1\rangle}}

\newcommand{\mred}[1]{{\color{red}{#1}}}

\newcommand{\im}{{\rm i}}

\begin{document}

\title{Closing the ``Quantum Supremacy'' Gap: Achieving Real-Time Simulation of a Random Quantum Circuit Using a New Sunway Supercomputer}

\author[1,3]{Yong (Alexander) Liu}
\author[1,3]{Xin (Lucy) Liu}
\author[1,3]{Fang (Nancy) Li}
\author[2,3]{Haohuan Fu}
\author[1,3]{Yuling Yang}
\author[1,3]{Jiawei Song}
\author[1,3]{Pengpeng Zhao}
\author[1,3]{Zhen Wang}
\author[1,3]{Dajia Peng}
\author[1,3]{Huarong Chen}
\author[4]{Chu Guo}
\author[4]{Heliang Huang}
\author[3]{Wenzhao Wu}
\author[2,3]{Dexun Chen}

\affil[1]{Zhejiang Lab, Hangzhou, China}
\affil[2]{Tsinghua University, Beijing, China}
\affil[3]{National Supercomputing Center in Wuxi, Wuxi, China}
\affil[4]{Shanghai Research Center for Quantum Sciences, Shanghai China}
\setcounter{Maxaffil}{0}
\renewcommand\Affilfont{\itshape\small}
\footnotetext[1]{For a finalized version, please refer to the published paper at https://doi.org/10.1145/3458817.3487399}

\date{\today}
\maketitle

\begin{abstract}
We develop a high-performance tensor-based simulator for random quantum circuits(RQCs) on the new Sunway supercomputer. Our major innovations include: (1) a near-optimal slicing scheme, and a path-optimization strategy that considers both complexity and compute density; (2) a three-level parallelization scheme that scales to about 42 million cores; (3) a fused permutation and multiplication design that improves the compute efficiency for a wide range of tensor contraction scenarios; and (4) a mixed-precision scheme to further improve the performance. Our simulator effectively expands the scope of simulatable RQCs to include the 10$\times$10(qubits)$\times$(1+40+1)(depth) circuit, with a sustained performance of 1.2 Eflops (single-precision), or 4.4 Eflops (mixed-precision)as a new milestone for classical simulation of quantum circuits; and reduces the simulation sampling time of Google Sycamore to 304 seconds, from the previously claimed 10,000 years. 
\end{abstract}

\medskip

A performance of 1.2 Eflops (single-preicision), or 4.4 Eflops (mixed-precision) for simulating a 10$\times$10$\times$(1+40+1) circuit (a new milestone for classical simulation of RQC), using about 42 million Sunway cores. The time to sample Goolge Sycamore in a simulation way is reduced from years to 304 seconds.

\section{Performance Attributes}

\vspace{-2mm}
\begin{table}[ht]
\centering
\begin{tabular}{ll}
\hline
\textbf{Performance Attributes}       & \textbf{Content}                                                                                                                                        \\ \hline
Perfomance                   & \makecell[l]{1.2 Eflops (single-precision) \\4.4 Eflops (mixed-precision)}                                                                               \\ \hline
Maximum problem Size & \makecell[l]{10$\times$10 (qubits) $\times (1+40+1)$ (depth)  }                                                                \\ \hline
Category of achievement      & \makecell[l]{Peak performance, and time to\\ solution}                                                                                              \\ \hline
Type of method used          & \makecell[l]{Simulating the RQC by \\ contracting a tensor network}                                                                                                                                           \\ \hline
Results reported on basis    & Whole application                                                                                                                              \\ \hline
Precision reported           & \makecell[l]{Single precision, \\and mixed precision (single/half)}                                                                                                                               \\ \hline
System scale                 & 107,520 nodes (41,932,800 cores)                                                                                                                  \\ \hline
Measurement mechanism        & Flops counts and timers                                                                                                                        \\ \hline
\end{tabular}
\end{table}

\section{Overview of the Problem}




With Google's ``Quantum Supremacy'' declaration in 2019 \cite{google-nature-2019}, stating that the Sycamore superconductive quantum computer is over a billion times faster than Summit \cite{osti_1259664} (comparing 200 seconds against 10,000 years in the task of measuring/simulating one million samples), the dawn of the quantum age starts to unfold in a more affirmative way. A later response from the IBM research team \cite{pednault2019leveraging} argues that they can accomplish the simulation on the classical Summit supercomputer, by using the hard disk to overcome the exponentially increasing complexity, within a few days instead of 10,000 years.

The progresses are promising and exciting on various fronts of quantum technology \cite{google-nature-2019,zhong2020quantum}. However, the notation of ``Quantum Supremacy'' is in many cases ``confusing'' and ``misleading''. For example, most researchers outside the quantum discipline would struggle a while to understand that instead of describing a general advatange, the word only refers to a specific quantum scenario that the quantum computing solution demonstrates superpolynomial speedup over the classical computing solution \cite{preskill2012quantum}. A typical example is the task to sample the quantum states of a Random Quantum Circuit (RQC) that entangles the different quantum bits, i.e. qubits. The quantum physics behind the entangling qubits requests the classical binary bits to store and compute the information with an exponentially-increasing complexity.  

Facing the ``quantum fuzz'', we think that it is extremely important to develop a highly efficient classical simulator that can utilize the emerging exascale systems to simulate some of the most recent quantum systems. Our main motivations are as follows:

\begin{itemize}
	\item First, a highly efficient quantum circuit simulator would provide concrete support for current development of noisy intermediate-scale quantum (NISQ) computing devices\cite{Preskill_2018,Wright_2019}. Even though the word ``quantum computing'' is already a hot scientific topic in public media, the technology itself is still at a starting stage, similar to the time when we just had electronic tubes invented as the basic building blocks of electronic computers. At the current stage, the technology for generating, manipulating, and measuring the qubits still involves a high level of uncertainty. The fidelity of the 2 million samples measured from the 53-qubit Sycamore\cite{google-nature-2019,zlokapa2020boundaries}, described in terms of the linear cross-entropy benchmark (XEB), is only at a level of 0.2\%. In contrast, the simulator using a classical computer, although working with an exponentially-increasing complexity, can provide significantly higher fidelity \cite{pan2021simulating}, and serve as an important guide and reference for the quantum computer design process.
	\item Second, for a leading-edge supercomputer, we think that taking a challenge that is inherently exponentially more complex could potentially bring algorithmic and architectural innovations within the traditional supercomputing community. There is an old saying from the Classic of Poetry (Shijing) of China, ``a stone from a remote mountain could serve to polish a local jade.'' The completely different way of computing in a quantum environment could potentially lead to a rethinking of algorithms for many cases on classical computers.
\end{itemize}

Based on the above considerations, we develop a highly efficient random quantum circuit (RQC) simulator for the new generation of Sunway Supercomputer. With a systematic design process that covers the algorithm, the parallelization, the architecture, and the precision aspects, we propose a complete approach that includes: (1) a near-optimal slicing scheme that achieves a good balance between compute and storage costs, and a multi-objective strategy that optimizes the path for both a reduced complexity and a balanced compute density; (2) a customized parallelization scheme that maps the tensor computation to about 42 million cores of the system; (3) a fused permutation and multiplication design that fully utilizes the machine's memory hierarchy and vectorization capacity, so as to support high-performance computation of different tensor contraction cases; (4) a mixed-precision algorithm that applies both single- and half-precision arithmetic to achieve a same level of accuracy through an adaptive scaling. With our RQC simulator on Sunway, we are able to simulate a 2D array of 10$\times$10 qubits with a depth of $(1+40+1)$. The simulator achieves a sustained performance of 1.2 Eflops (single-preicision), or 4.4 Eflops (mixed-precision), demonstrating a highly efficient utilization of the new Sunway system. In contrast to Google Sycamore, which produces one million samples with a fidelity of 0.2\% in 200 seconds, we can accomplish the sampling, with a strict `Quantum Supremacy' comparison configuration, in a real-time scale of 304 seconds.  

To the best of our knowledge, the 10$\times$10$\times$(1+40+1) RQC is the largest RQC that gets simulated using a classic supercomputer. The sampling of Google Sycamore in 304 seconds is also the first time that we are closing the simulation time to the scale of seconds, instead of days, or even thousands of years claimed in previous ``quantum supremacy'' comparisons.

\section{Current State of the Art}

\subsection{General Background of RQC}

With the goal to accomplish computation tasks, the quantum computers are built with the quantum bits (qubits) as the basic block. A qubit is usually described as a linear combination of two orthonormal basis states $\ket{0}$ and $\ket{1}$ (written in the traditional Dirac notation), shown as follows:
\begin{equation}
\ket{\psi} = \alpha\ket{0} + \beta\ket{1},
\end{equation}
where $\alpha^2+\beta^2=1$. Compared with the classical bit being either 0 or 1, the qubit is considered as a coherent superposition of the both the $\ket{0}$ state and the $\ket{1}$ state. $\alpha$ and $\beta$ describe the probability of the qubit, when measured, collapsing to the $\ket{0}$ state (with a probability of $\alpha^2$), or the $\ket{0}$ state (with a probability of $\beta^2$). 

\begin{figure}[ht]                                    
    \centering 
    \includegraphics[width=0.6\textwidth]{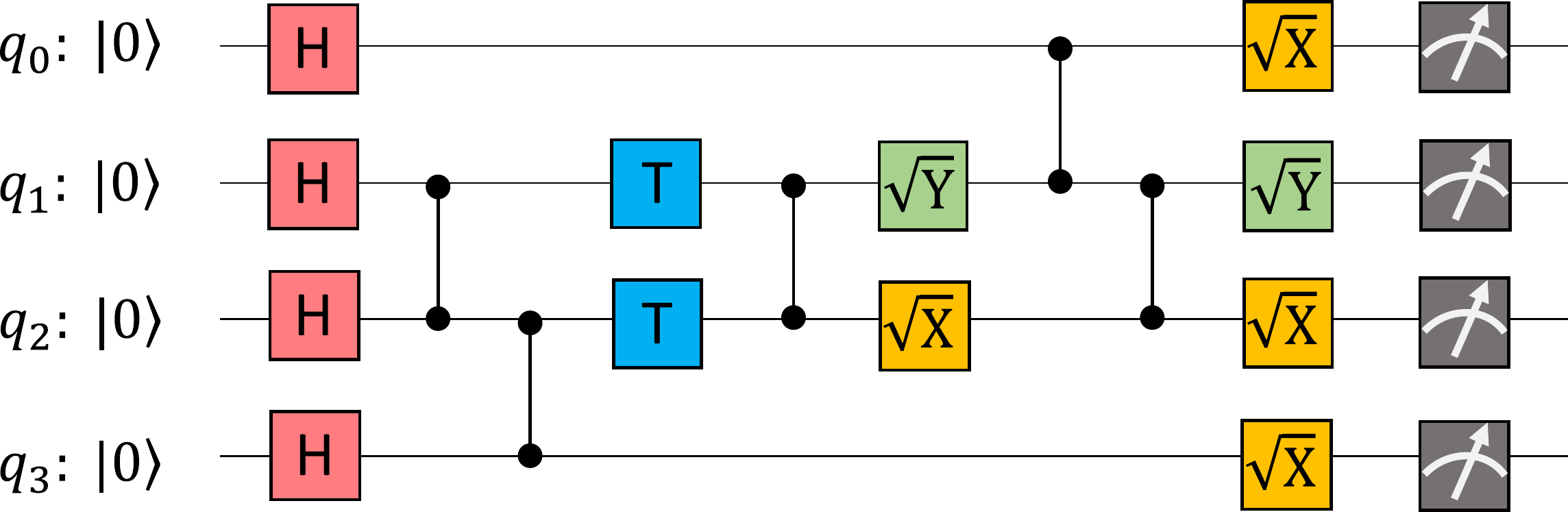}
    \caption{An example diagram of a typical random quantum circuit.}
    \label{fig:rqc-example}   
\end{figure}

Building on top of the qubits, a typical RQC (as shown in Fig. \ref{fig:rqc-example}) would involve a random sequence of gates to manipulate the states of the qubits. The number of different levels of gates in between the input qubits and the output qubits are described as the depth ($D$) of the RQC (in a quantum system, the levels usually correspond to manipulations at different cycles). 

Going through the gates in between the input and output, the multi-qubit quantum state becomes highly entangled, which is another unique feature of quantum mechanics, leading to the exponetial advantage of the quantum computing system. In an entangled quantum system, to describe the states of the $n$ qubits, we need a space of $O(2^n)$, which may easily exceed the memory space of current leading-edge supercomputers (a 49-qubit system requires 8 PB in double precision), and a similar level of computational complexity that makes the classical simulation significantly slower than quantum.    

\subsection{Existing Efforts on Simulation of RQCs}

With the fast development of NISQ devices in recent decades, different classes of RQC simulators on classical computers have blossomed to tackle the demanding problem of simulating large quantum circuits. To date, simulators of quantum circuits can be classified into the following two main categories.

\textbf{\emph{(1) The state vector approach that simulates a direct evolution of the quantum state.}}

To maintain a general capability to simulate an arbitrary form of RQC, the ``Schr{\"o}ndinger'' kind of approach usually stores a full state of the qubits. The exponential increase of the memory space needed ($O(2^n)$ space for an $n$-qubit system) becomes the biggest challenge. An early work in 2007 \cite{de2007massively-11yearsbefore} utilizes 1 TB of memory on some of the largest supercomputers at that time (IBM BlueGene/L, Cray X1E, Hitachi SR11000/J1) to simulate a 36-qubit system. A more recent effort on Cori II \cite{sc17-half-petabyte-simulation-45qubit} uses 0.5 PB of memory to simulate a quantum circuit with 5$\times$9=45 qubits of depth 25, with the increase of the memory capacity and the circuit complexity strictly following the $O(2^n)$ pattern. 

Along the way, various techniques were proposed to reduce the memory requirement, so as to enable simulation of larger quantum circuits. For example, as an eleven-year-later update to the aforementioned 2007 work \cite{de2007massively-11yearsbefore}, adaptive encoding of the wave function was used to reduce the memory requirement by a factor of 8, and enabled simulation of 48 qubits using 0.5PB of memory \cite{11yearslater}. A recent work\cite{li-sunway-2019quantum} on Sunway TaihuLight\cite{fu2016sunway}  targets the universal random quantum circuits. By taking advantage of the diagonal properties of CZ gates, an implicit decomposition scheme is proposed to accomodate the problem of simulating 7$\times$7 qubits of depth 39, and can achieve the result of all amplitudes in 4.2 hours. Moreover, lossless compression and lossy compression with adaptive error bounds \cite{wu-sc2019-full} were also adopted to reduce the memory requirement of a 61-qubit system from 32 EBs to 768 TBs, making it a possible simulation on Argonne's Theta supercomputer.

\textbf{\emph{(2) The tensor approach that performs the simulation by contracting a tensor network.}}

In the tensor type of approach, the quantum circuit is described as a tensor network, where a one-qubit gate is described as a rank-2 tensor, and a two-qubit gate is described as a rank-4 tensor, and an $n$-qubit gate is described as a rank-2n tensor \cite{biamonte2017tensor-nutshell}. The simulation of the quantum circuit is then transformed into a problem of contracting the corresponding tensor network, where we perform the convolution of the corresponding tensors until there is only one vertex left.

A straightforward implementation of the tensor approach would in general involve a complexity that also grows exponentially with the number of qubits and the depth of the circuit, making simulation of large-scale circuits difficult to accomplish within a reasonable amount of time. However, if we perform the simulation by doing only one or a small batch of the state amplitudes at the end of circuit, the complexity of the tensor network would be constrained by the largest tensor involved in the contraction process, which grows exponentially with the tree width of the graph corresponding to the tensor network \cite{contracting-tensor}. 

Such a method can be highly efficient for circuits with a large number of qubits but a shallow depth, as complexity also grows exponentially with the depth of the circuit in many cases. For example, the undirected graphical model proposed by the Google research team \cite{google-unigraph} uses a workstation to sample 7$\times$8 qubits of depth (1+30+1) in an efficient way (with each sample taken in roughly 600 seconds), but the 7$\times$7 qubits of depth (1+40+1) was still out of reach. 

A later work from Alibaba \cite{chen2018classical} then further pushes the frontier to simulation of 8$\times$8 qubtis of depth (1+40+1) (using a similar undirected graphical model), with each amplitude simulated within 2 minutes, using around 131,000 compute nodes in the AliCloud. However, it was argued that the performance only holds for a specific circuit class \cite{villalonga2020establishing}, and would be at least 1,000 times easier than the circuits discussed in \cite{ai-blog}.  

Another technique that can be integrated into the tensor-based approach is the Projected Entangled-Pair States (PEPS) representation of quantum states from many-body quantum physics. By using PEPS as the data structure to represent the quantum states, a general purpose quantum circuit simulator is built \cite{PEPS-GUO-2019} to simulate 8$\times$8 qubits of depth 25 on a personal computer, and 10$\times$10 qubits of depth 26 on the Tianhe-2 Supercomputer.

Based on the previous efforts on the undirected graphical model of Google \cite{google-unigraph}, a collaboration between Google and NASA presented qFlex \cite{google-simulator-2018}, a flexible tensor network based simulator for quantum circuits, with a focus on RQCs in the range of sizes expected for ``quantum supremacy'' experiments. With considerations for both the capability to simulate an arbitrary RQC circuit and the efficiency on a large-scale supercomputer, the code supported a 60-qubit lattice simulation that used around 90\% of the nodes NASA HPC clusters Pleiades and Electra, sustained a peak performance of 20 Pflops with the two clusters combined. A recent following effort further mapped qFlex to Summit, one of the fastest supercomputers in the world\cite{villalonga2020establishing}, and took advantage of a rich set of GPU accelerators. The GPU version of qFlex achieved a sustained performance of 281 Pflops when using the entire Summit system to simulate 7$\times$7 qubits of depth (1+40+1).
\begin{figure*}[htbp]                               
    \centering 
    \includegraphics[scale=0.32]{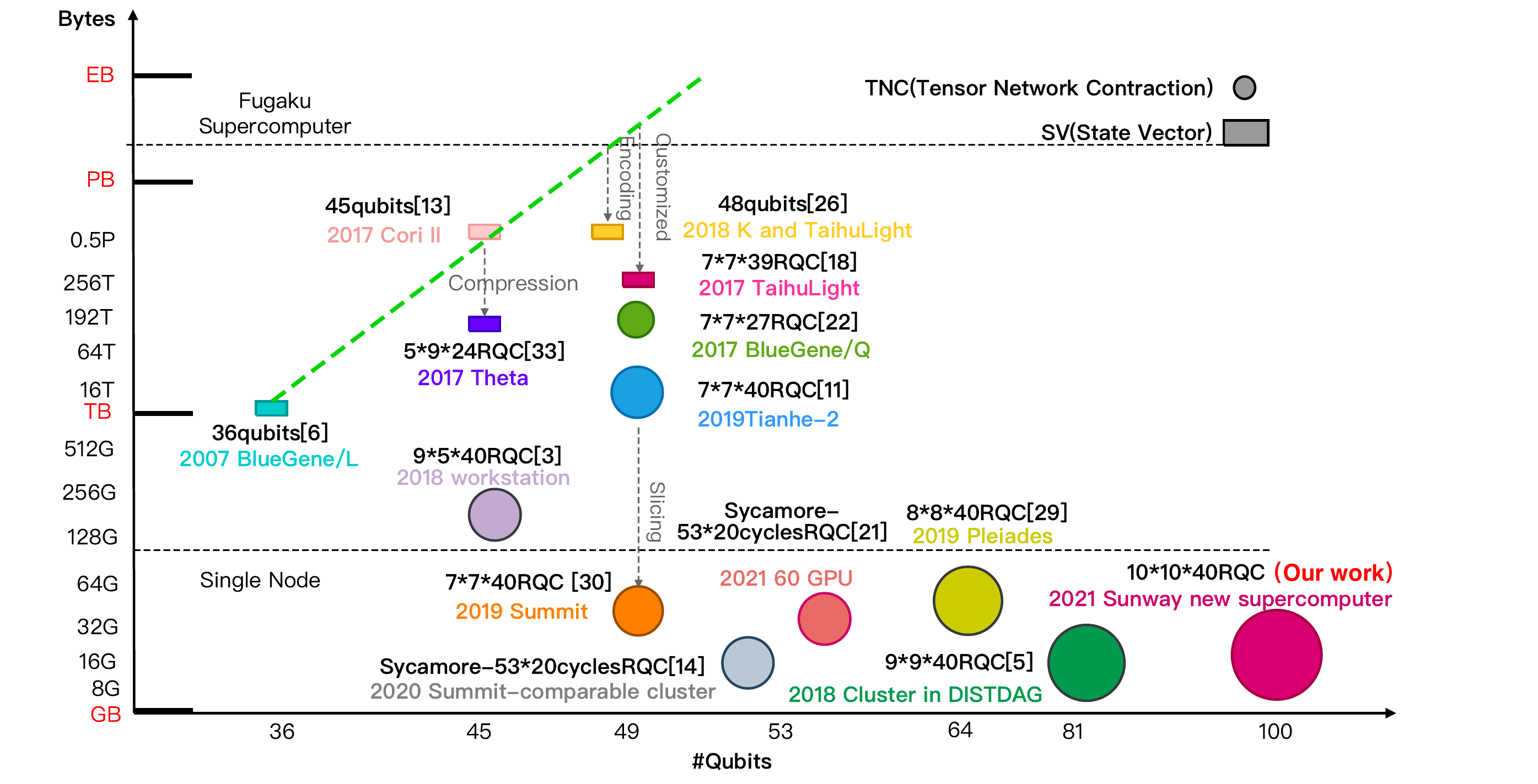}
    \caption{A summary of major classical RQC simulations. The x-axis denotes the number of qubits, while the y-axis shows the corresponding memory space required. The size of the circle/rectangular corresponds to the complexity (depth) of the circuit.}
    \label{fig:rqc-memory-compute}   
\end{figure*}

\subsection{Comparing Classical Simulators with the Google Sycamore System}

With the Google Sycamore system demonstrating the ``Quantum Supremacy''\cite{google-nature-2019} for the first time, a number of recent RQC simulation efforts target the Sycamore system for an exact comparison.

A direct reply from the Summit system \cite{pednault2019leveraging}, after the announcement of Sycamore, claimed that by using the hard disk storage of Summit, the team could accomodate the 53-qubit and 54-qubit scenario (using the state vector approach) and finish the simulation in a few days, instead of 10,000 years stated in \cite{google-nature-2019}. While this article provides a quick response to the Sycamore paper, it is still at a stage of ideas rather than results.  

A more concrete work is achieved by the Alibaba Quantum Lab, using a Summit-comparable cluster\cite{huang2020classical}. By testing the run time of all the basic subtasks, it is estimated that the Sycamore measurement task (20 cycles circuit with a 0.2\% fidelity) can be accomplished within 20 days, using a tensor-based approach and a specific strategy to identify and to optimize the `stem', i.e. the main path for contracting the tensor pairs.

A recent work \cite{Schutski_2020} propose a parallel algorithm for the contraction of tensor networks using probabilistic graphical models. Another recent work \cite{guo2021verifying} present a tensor network states based algorithm specifically designed to compute amplitudes for random quantum circuits with arbitrary geometry.

The latest work \cite{pan2021simulating} achieves a computational complexity that is much lower than both the Google simulation in the ``Quantum Supremacy'' statement \cite{google-nature-2019} and the Alibaba work \cite{huang2020classical}, by using a subspace sampling technique instead of random sampling in the full Hilbert space. Using 60 GPUs, the proposed method could already compute exact amplitudes of 2 million correlated bitstrings in 5 days.

\subsection{Summary}

Fig. \ref{fig:rqc-memory-compute} shows a general picture about the development of all different classical simulation methods over the years. As the time complexity is very implementation-specific, we put the space complexity (corresponding to the total memory size) as a major metric to demonstrate the evolution of different types of methods. 

The state vector type of methods follow a strict pattern of an $O(2^{n})$ space complexity for an $n$-qubit system, shown as the green dotted line (examples such as \cite{de2007massively-11yearsbefore,sc17-half-petabyte-simulation-45qubit} sit exactly on the line). Various techniques were proposed to divert slightly from the line, such as the compression method in \cite{wu-sc2019-full}, the encoding method in \cite{11yearslater}, or customization for a specific circuit in \cite{li-sunway-2019quantum}. Even with the diversion provided by different techniques, the steep slope would soon touch the upperbound of Fugaku \cite{JDongarraReportFugaku}, the supercomputer with the largest memory space on the current top500 list. 

In contrast to the state vector type of methods, the tensor contraction methods can significantly reduce the required memory space\cite{villalonga2020establishing,google-simulator-2018,huang2020classical}. Especially the slicing method (discussed in more details in Section \ref{sec:sim-method}) would reduce the required memory space from PB to TB or even GB scale. As a result, most of recent efforts\cite{pan2021simulating,chen2018classical,villalonga2020establishing,google-simulator-2018,huang2020classical} use such a strategy to fit the tensor contraction computation into a single node, and scale the performance in a highly-efficient way by allocating different tensors to different computing nodes.

Our simulation also takes the tensor-based approach. In contrast to the other tensor methods, we apply a heuristic method to identify the most suitable slicing scheme, as well as the resulting contraction order. By adopting such strategies on the new Sunway supercomputer, we can simulate complex RQCs with $10\times10$ (with a depth of (1+40+1)) or even $20\times20$ qubits (with a depth of (1+16+1)). As far as we know, this is the largest quantum circuit that gets simulated on a classical supercomputer. 

Compared with the Google Sycamore system \cite{google-nature-2019}, which declared the ``Quantum Supremacy'' in 2019 with one million samples generated in 200 seconds with a fidelity of 0.2\%, our simulator can provide similar samples within seconds instead of years, thus capable of providing real-time simulation for current quantum systems.

\section{The New-Generation Sunway Supercomputer}
\subsection{System Architecture}
\begin{figure}[ht]                                    
    \centering 
    \includegraphics[scale=0.27]{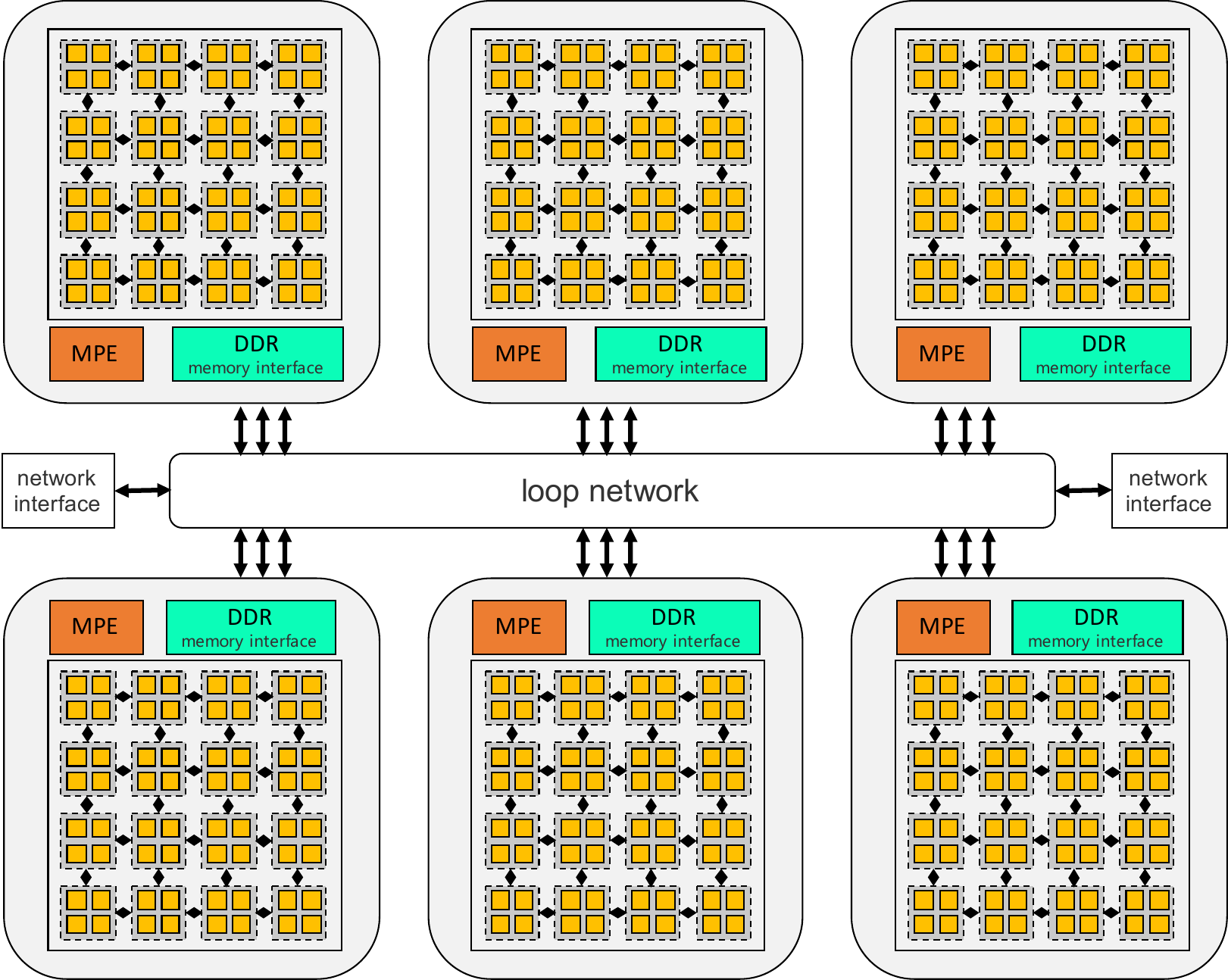}
    \caption{General architecture of the SW26010P CPU.}
    \label{fig:sw390}   
\end{figure}

The computing capability of the new-generation Sunway supercomputer is provided by a homegrown many-core SW26010P CPU that includes 6 core-groups (CGs), each of which includes one management processing element (MPE), and one 8$\times$8 computing processing element (CPE) cluster, as shown in Fig. \ref{fig:sw390}. 

Each CG has its own memory controller (MC), connecting to 16 GB of DDR4 memory with a bandwidth of 51.2 GB/s. The data exchange between each two CPEs in the same CPE cluster is achieved through the Remote Memory Access (RMA) interface (a replacement of the register communication feature in the previous generation). Each CPE has a fast local data memory (LDM) of 256 KB.

Each SW26010P processor consists of 390 processing elements. Our largest experiment uses a total of 107,520 SW26010P CPUs, with an unprecedented parallel scale of 41,932,800 cores. While the new system demonstrates significant advantages over other systems on the computing capability, similar to its predecessor, the system is relatively modest in terms of the memory capacity, with 96 GB memory and a memory bandwidth of 307.2 GB/s for each node.

\subsection{Simulating RQCs on Sunway: Major Design Challenges}
\label{sec:challenge}


The goal that we try to achieve here, is to simulate the world's leading RQCs (a delibrately designed quantum scenario that is expected to take thousands of years to compute) on the world's leading supercomputers (with tens of millions of cores). In other words, we need to map a classical-computer-unfriendly problem to the world's largest classical computer.

Therefore, the algorithm-wise challenge is to, based on all the existing RQC simulation algorithm efforts, explore and identify the one algorithm that provides the best mapping of the required computation to the underlying supercomputer, in terms of parallelism, compute and memory complexity, as well as the compute to memory ratio (a key factor to achieve fair performance on today's heterogeneous many-core supercomputers). 

With the strict memory requirement of the state vector approach that constrains the supported size of RQCs, we take the tensor-based approach \cite{contracting-tensor}, which is also the category with more active efforts in recent years. Our starting point is the PEPS method \cite{PEPS-GUO-2019}, with a heurisitc method to identify a contraction path that can effectively reduce the order of largest intermediate tensors. Following that, we further adopt the hyper-optimized search of the CoTenGra software \cite{gray2021hyper} to explore and identify the most suitable `stem' that provides enough parallelism, significantly reduced compute cost, as well as a reasonable flop-to-byte ratio that keeps the cores busy enough.

To support such an algorithm design on a Sunway supercomputer, the implementation-wise challenge can be manyfold:
\begin{itemize}
\item To parallelize the computation efficiently over 40 million cores, we need a scheme that, at the first level, can efficiently decompose the tensor contraction into different sub-tasks corresponding to different MPI processes; and at the second level (a typical sub-task would involve two CGs, with a memory capacity of 32 GB and a peak performance of 4.7 Tflops), can  further distribute the tensor operations to the 130 processing elements. In Section \ref{sec:parallel-scheme}, we discuss our method to maintain the balance between storage and computational complexity at the level of the CG pair.
\item To achieve compute efficiency of the major tensor kernels, we need a generalized tuning strategy to map a wide range of tensor kernels (different tensor contraction cases, produced by different contraction `stems' in our path exploration process) to the underlying 8 by 8 CPE mesh, and the wide vector units in each CPE. 
\item Facing such a problem with an extreme level of computational complexity, the potential of mixed-precision arithmetic, supported by most exa-scale systems nowadays, becomes a necessary component to further boost the performance. Section \ref{sec:mixed} introduces our early efforts on investigating the potential of mixed-precision methods in simulating RQCs (improving the sustained performance from 1.2 Eflops to 4.4 Eflops).
\end{itemize}

\section{Innovations}

Although the tensor-based approach avoids the hard constraint for a large memory space to store all possible quantum states, the involved compute cost itself is already an Alps-level mountain to climb. For the $10\times10\times(1+40+1)$ RQC that we target in this work, a straightforward implementation of the original tensor approach \cite{contracting-tensor} involves a computational complexity at the level of $10^{10}$ Eflops, requiring the new generation Sunway supercomputer to compute continuously for 7,610 years (assuming 100\% compute efficiency). 

Therefore, the first step to accomplish our target simulation is to find a method to reduce the complexity to a level that is practical for the current supercomputers. Thanks to the continuous efforts by different researchers in recent years \cite{PEPS-GUO-2019,google-simulator-2018,huang2020classical,pan2021simulating}, we have a number of techniques for reducing the computational complexity. Section \ref{sec:sim-method} presents our PEPS-based simulation scheme, with a heurisitic method to determine an optimized slicing scheme that fits well to both the parallel scale and the underlying many-core architecture. While the PEPS-based approach brings the best performance for the 10$\times$10$\times$(1+40+1) RQC, the Sycamore circuit is still out of its reach (due to the depth brought by the $fsim$ gates). Section \ref{sec:3d-path} describes our further efforts to search and to identify the most suitable contraction path that further reduces the complexity in a significant manner, and finally to make a strict Quantum Supremacy simulation of Sycamore feasible on the Sunway machine.

As mentioned in Section \ref{sec:challenge}, the algorithmic efforts bring further requirements for parallelization, kernel tuning, and adoption of mixed-precision arithmetic. Sections \ref{sec:parallel-scheme}, \ref{sec:tensor-op}, and \ref{sec:mixed} describe our efforts in those aspects respectively. 

\subsection{A PEPS-based Simulation Method with an Optimized Slicing Strategy}
\label{sec:sim-method}

Our first tensor-based method adopts the PEPS approach, which uses the projected entangled-pair states (PEPS) to represent the tensor-network quantum states for 2D lattices. Taking the PEPS approach, the computational complexity of the RQC simulation is determined by the underlying quantum circuit rather than the number of qubits or gate operations. In our target case of the 10$\times$10$\times$(1+40+1) RQC, the complexity is in the range of $2^{76}\approx7558$ Eflops. With a high efficiency of the simulation software, such a complexity is already within the reach of current exa-scale systems.

\begin{figure}[ht]                                    
    \centering 
    \includegraphics[scale=0.4]{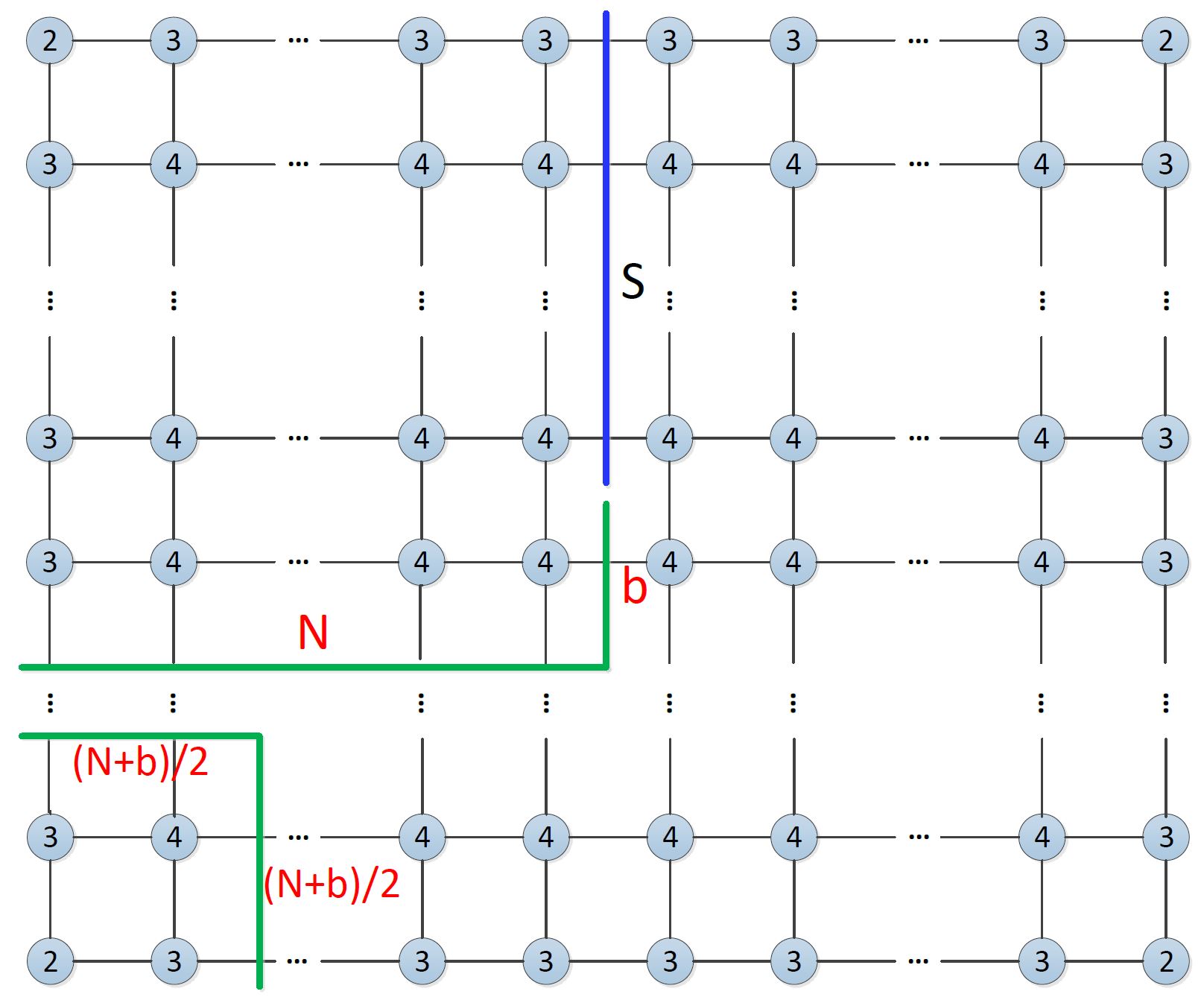}
    \caption{Our optimized slicing scheme for a rectangular tensor network.  The blue line represents the position of slicing action, which cuts $S$ hyperedges. $S={3(N-b)}/2$, $b=2-\delta_{odd}(N)$. $\delta_{odd}(N)=1\quad(if\quad N \quad is \quad odd)\quad or \quad 0 \quad (if \quad N \quad is \quad even)$. The green line represents the main tensors contraction in each contract path. Before the slicing, the fewest space complexity is $O(L^{2N})$, with $O(L^{3N})$ computing complexity, where $L=2^{\lceil d/8\rceil}$, and $d$ is the depth of the RQC. After the slicing, the space complexity is reduced to $O(L^{N+b})$, while the time complexity remains at the scale of $O(L^{3N})$.}
    \label{fig:tensor-slicing}   
\end{figure}

In tensor-based approaches, slicing is a commonly used technique to balance the memory requirement and the number of concurrent computations that we can perform \cite{breaking-49-qubit}. For example, for the 2n$\times$2n qubit lattice shown in Fig. \ref{fig:tensor-slicing}, the blue line cuts off the $S$ hyperedges of the top $S$ rows. For all the possible $2^S$ values of the $S$ hyperedges, we can perform the contraction of the $2^S$ sliced tensors with the hyperedges fixed to one of the $2^S$ possible values. With the slicing approach taken, the memory requirement to store the tensor can be further reduced (usually at the order of $2^S$), and the contraction of the original network can be converted into $2^S$ independent contraction sub-tasks. Therefore, the slicing step generally becomes the natural scheme to perform the first level of task decomposition for a large-scale parallel computing environment \cite{villalonga2020establishing,huang2020classical}.

Finding an optimized slicing scheme is important for achieving an efficient simulation of RQC on a supercomputer. As the slicing brings both reduced space complexity of the sliced tensor and embarrassing parallelism of the subtasks, we need to identify a balance point with subproblems that fit well into the memory space of each MPI process and an acceptable increase in the compute cost aggregated from all the parallel subtasks.

Targeting a $2N\times2N$ rectangular tensor network, we propose a simple and straightforward heuristic method to identify the near-optimal slicing scheme and the contraction order (as shown in Fig. \ref{fig:tensor-slicing}). For the entire contraction process, we keep the contraint that the rank of the tensor should not exceed $N+b$. The value of $b$ is determined by the parity of $N$ (we set $b$ to be 1 if $N$ is odd, or 2 if $N$ is even). With such an strategy, the $(N+b)/2$ by $(N+b)/2$ square tensor network can be contracted from the lower-left corner, resulting in a tensor with a rank of $N+b$. Denoted as the blue line in Fig. \ref{fig:tensor-slicing}, we perform the slicing of $S$ hyperedges, and keep the rank of the tensor that contracts downward to be also within $N+b$. These two tensors have $(N+b)/2$ connecting hyperedges, leading to a contraction complexity of $L^{(3(N+b)/2)}$, where $L=2^{\lceil d/8 \rceil}$, and $d$ is the depth of the circuit. 

To perform such a slicing, as shown in Fig. \ref{fig:tensor-slicing}, the number of sliced hyperhedges $S$ can be calculated as $S=2N-(N+b)/2-b=3(N-b)/2$.

Following such a slicing strategy, we can find the near-optimal contraction order, which keeps the computational complexity at the scale of $O(2\cdot L^{(S+3(N+b)/2)})=O(2\cdot L^{3N})$. As noted in \cite{PEPS-GUO-2019}, such a computational complexity is similar to the time complexity of a minimized space complexity without slicing. Therefore, our slicing method, while significantly reducing the memory requirement of a classical simulation, maintains a computational complexity that is close to optimal. As shown in Fig. ~\ref{fig:tensor-slicing}, such a strategy can effectively support a larger size of the circuit, and enables a more balanced configuration between communication and compute. 

As required by the frugal rejection sampling technique \cite{google-simulator-2018}, while $10^6$ sampled amplitudes are enough for cross-entropy verification of the RQC sampling task, we often need to simulate 10 times more ($10^7$) amplitudes for correct sampling on a classical supercomputer. To reduce the sampling cost, we follow the idea of fast sampling technique \cite{google-simulator-2018} to select a number of qubits as the open batch. For the 10$\times$10 qubit lattice example, we compute 512 amplitudes in a batch, with an overhead of only 0.01\% when compared with the normal approach of computing a single amplitude.

\subsection{Searching for Better Contraction Paths with a Reduced Complexity}
\label{sec:3d-path}

While our PEPS-based approach described in Section \ref{sec:sim-method} focuses on 2N by 2N lattice-type RQCs, we also support a generalized tensor-based simulation of RQCs in other shapes, such as Google Sycamore.  

\begin{figure}[ht]
\centering
\subcaptionbox{Sycamore}{\includegraphics[width=0.25\textwidth]{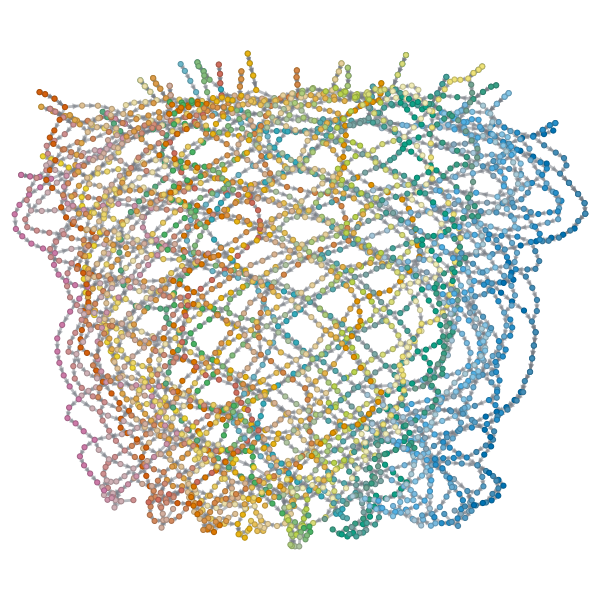}}%
\hfill
\subcaptionbox{10x10x(1+40+1)}{\includegraphics[width=0.25\textwidth]{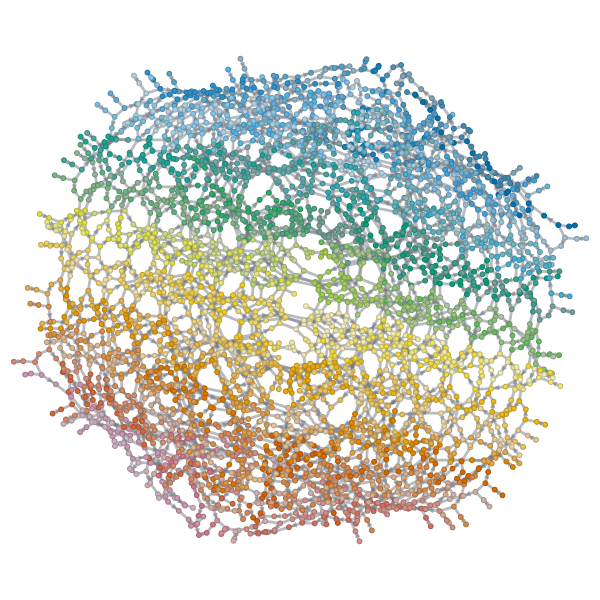}}%
\hfill
\subcaptionbox{Zuchongzhi}{\includegraphics[width=0.25\textwidth]{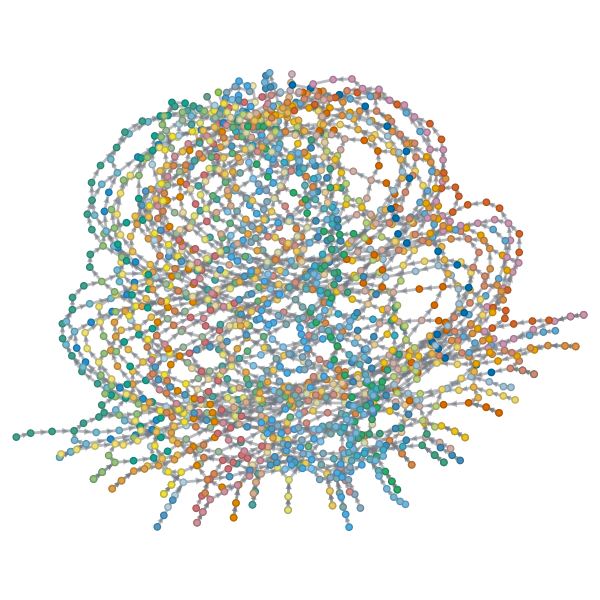}}%
\caption{The example tensor network of typical RQCs.}
\label{fig:rqc-graph} 
\end{figure}

In tensor-based methods, the problem of simulating the output of a specific sample or a batch of samples becomes the contraction of the corresponding tensor network. Fig. \ref{fig:rqc-graph} shows example tensor networks for typical RQCs, such as Sycamore \cite{google-nature-2019}, Zuchongzhi-One \cite{gong2021quantum}, and our target $10\times10\times(1+40+1)$ RQC. For such complex uni-directional graphs, different contraction paths lead to computational complexities that could differ by orders of magnitude. Therefore, finding a best contraction path becomes a central problem \cite{gray2021hyper}. 


In a context of mapping the simulation to a exa-scale supercomputer like Sunway, the definition of the `best' contraction path becomes a multi-objective issue. While minimizing the computational complexity is important, generating tensor ranks that would better fit the underlying many-core architecture is also crucial for achieving efficiency and simulation speed. To tackle such a complex situation, we apply the CoTenGra software \cite{gray2021hyper} to search for the best path with a loss function that combines the considerations for both the computational complexity and the compute density, which can largely decide its performance on a many-core processor. 
\begin{figure}[ht]                                    
    \centering 
    \includegraphics[width=0.68\textwidth]{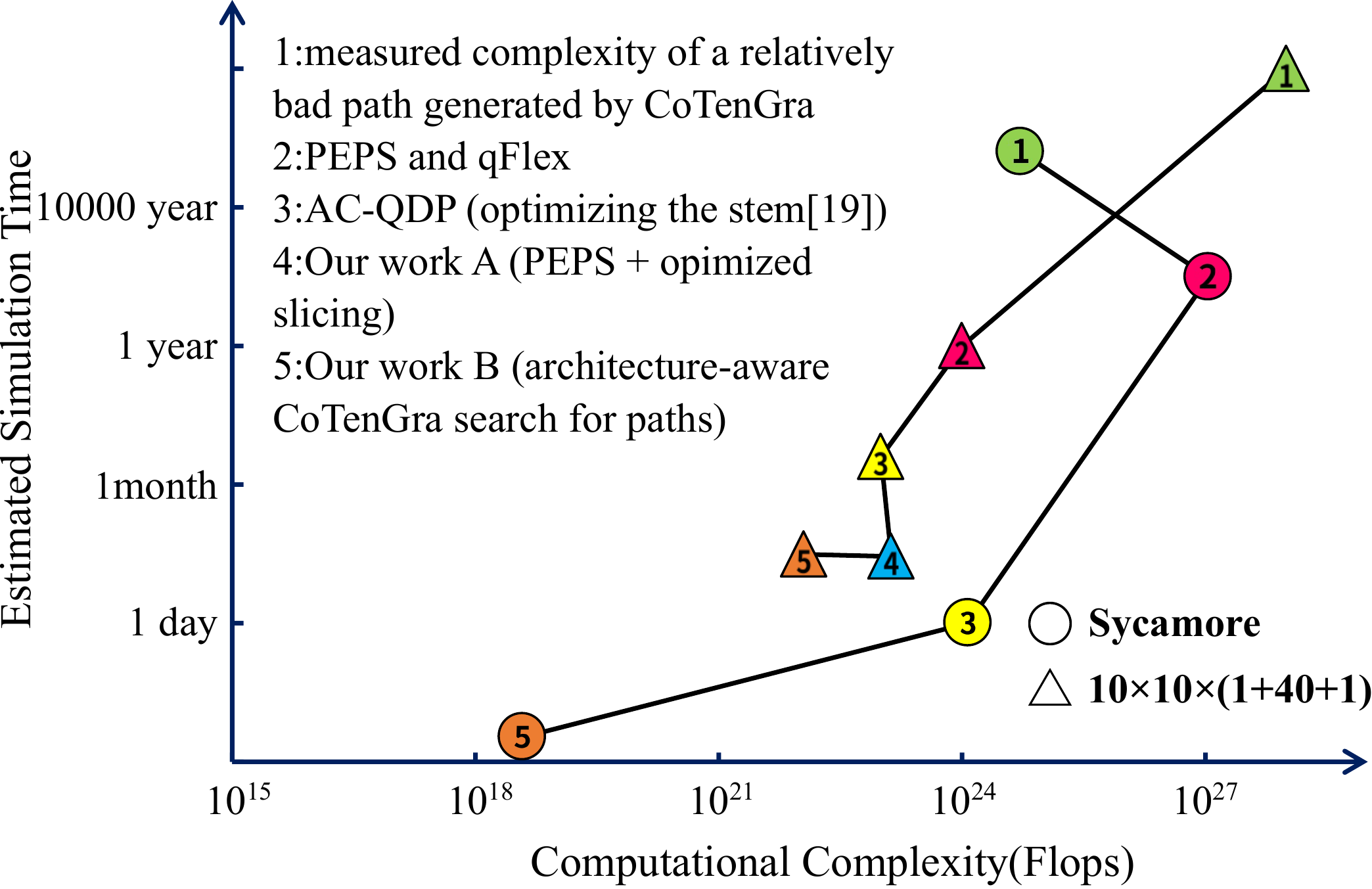}
    \caption{The computational complexities and their corresponding sampling time (measured or projected) for simulating the 10$\times$10$\times$(1+40+1) RQC and Google Sycamore, with different approaches to reduce the complexities. } 
    \label{fig:complexity} 
\end{figure}

Fig. \ref{fig:complexity} shows the computational complexities and their corresponding sampling time (measured or projected) for simulating the 10$\times$10$\times$(1+40+1) RQC and Google Sycamore, with different approaches to reduce the complexities. The starting point for both circuits is a worst-case complexity selected from a number of unoptimized CoTenGra generated paths, indicating a baseline complexity we face for such kind of circuits. While the PEPS-based approach generally involves a middle-level of complexity, the 2D lattice compaction usually generate pair-wise tensor contractions with ranks around 5 or 6, and a dimension size of 32, making the tensor operations highly efficient on the many-core processors. Such a strategy provides the best performance as well as the best time to solution for the 10$\times$10$\times$(1+40+1) RQC, even though the computational complexity of the PEPS-based approach might be 10 times more than the best search result of CoTenGra. 

However, for the case of Sycamore simulation, with the $fsim$ gate making the increase of the complexity to an even higher level, a straightforward PEPS-oriented implementation of Sycamore would even increase the complexity to a level that is infeasible for the current Sunway system. The good part is that although the $fsim$ gates double the depth, the numbers of qubits and gates are still at a reasonable level (compared to the 10$\times$10$\times$(1+40+1) RQC). As a result, the CoTenGra produces a significantly better-optimized path on the Google Sycamore circuit compared to the 10$\times$10$\times$(1+40+1) RQC, showing a reduction in complexity by around a million times, compared to only ten times in the latter case.

Although the complexity issue is resolved, we now face both many-core-friendly contractions with a high compute density, and many-core-unfriendly contractions between high-rank tensors and low-rank tensors (typical tensor pairs in CoTenGra's genereated path for Sycamore), which demonstrate imbalance between compute and memory opeartions, and bring the challenges that we need to resolve in Section \ref{sec:tensor-op}.

\subsection{A Customized Parallelization Scheme}
\label{sec:parallel-scheme}

When simulating a 100-qubit circuit of depth (1+40+1), the computational space and time complexity becomes huge, and we need to decompose the problem into subtasks that run on different compute nodes of the supercomputer. 

\begin{figure*}[ht]                                    
    \centering 
    \includegraphics[width=0.95\textwidth]{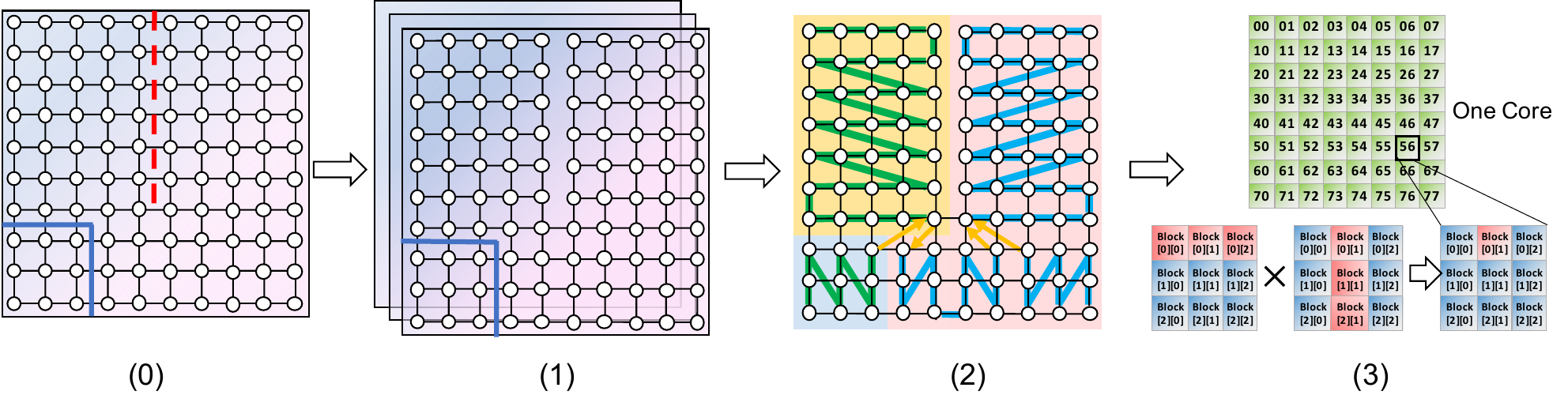}
    \caption{Our customized three-level parallelization scheme. The first level (0, and 1) uses the slicing method to decompose the 2D tensor network into a large number of independent sliced tensors, assigned to different MPI processes. The second level (2) decomposes the task to the two CGs. The third level (3) further decomposes the tensor permutation and matrix multiplication to different CPEs.}
    \label{fig:para}   
\end{figure*}

Following the slicing scheme mentioned above (as shown in Fig. \ref{fig:para}(0)), we first decompose the contraction of a 2D tensor network into independent tasks by cutting a number of hyperedges. In such a way, the computation of each amplitude of the quantum circuit can be divided into $L^S$ (here, $L=32$, $S=6$, derived from the optimized slicing scheme shown in Fig. \ref{fig:tensor-slicing}) subtasks. Each subtask, corresponding to the contraction of a sliced tensor, is allocated to a single MPI process, as shown in Fig. \ref{fig:para}(1). 

In our work, we represent each amplitude with two single-precision floating-point numbers (eight bytes). Therefore, the maximum space needed to store a sliced tensor is larger than $L^{(N+b)}\times8B=16GB$ ($N=5$, $b=1$), which is already touching the upper bound of  the total memory space of single CG. Considering the extra overhead of additional variables and system utilities, we adopt a CG pair to process the subtask of an MPI process. The partition between the two CGs is shown in Fig. \ref{fig:para}(2). The green and blue lines correspond to the tasks assigned to the two CGs respectively. After the contractions of green and blue parts are finished, the two CGs collaborate to process the contraction of the tensor with the largest rank (shown as the yellow lines in Fig. \ref{fig:para}(2)).

For the last level of parallelization among different CPEs, as shown in Fig. \ref{fig:para}(3), the specific matrix multiplication and index permutation are then mapped into efficient collaborative operations among different threads in different CPEs, detailed in Section \ref{sec:tensor-op}.

\subsection{Tensor Contraction Using Fused Permutation and Multiplication}
\label{sec:tensor-op}

The contractions of tensors have two basic steps: one is the index permutation of the tensors as a preparatory step; the other one is the following matrix multiplication to achieve the contraction result. 

The permutation of indices is generally required to convert the tensor contractions into efficient matrix multiplications. The permutation of the indices of high-rank tensors, which requires movements of data items with strides in between, is inherently unfriendly for current memory systems. As a result, to reduce or hide the permutation cost becomes a major design issue for achieving efficient tensor contractions on many-core processors with a high compute density \cite{villalonga2020establishing}.

For a rank-$m$ tensor $A(i_1, i_2, \cdots, k, \cdots, l, \cdots, i_m)$ and a rank-$n$ tensor $B(i_1, i_2, \cdots, k, \cdots, l, \cdots, i_n)$, assume that we need to contract on the $k$ and $l$ indices. In a normal implementation that separates permutation and multiplication, if $k$ and $l$ are neighboring indices, then we only need to perform the permutation for once. Otherwise, we may need to perform the permutation for multiple times, leading to significant permutation overhead.

\begin{figure}[ht]                                    
    \centering 
    \includegraphics[width=0.75\textwidth]{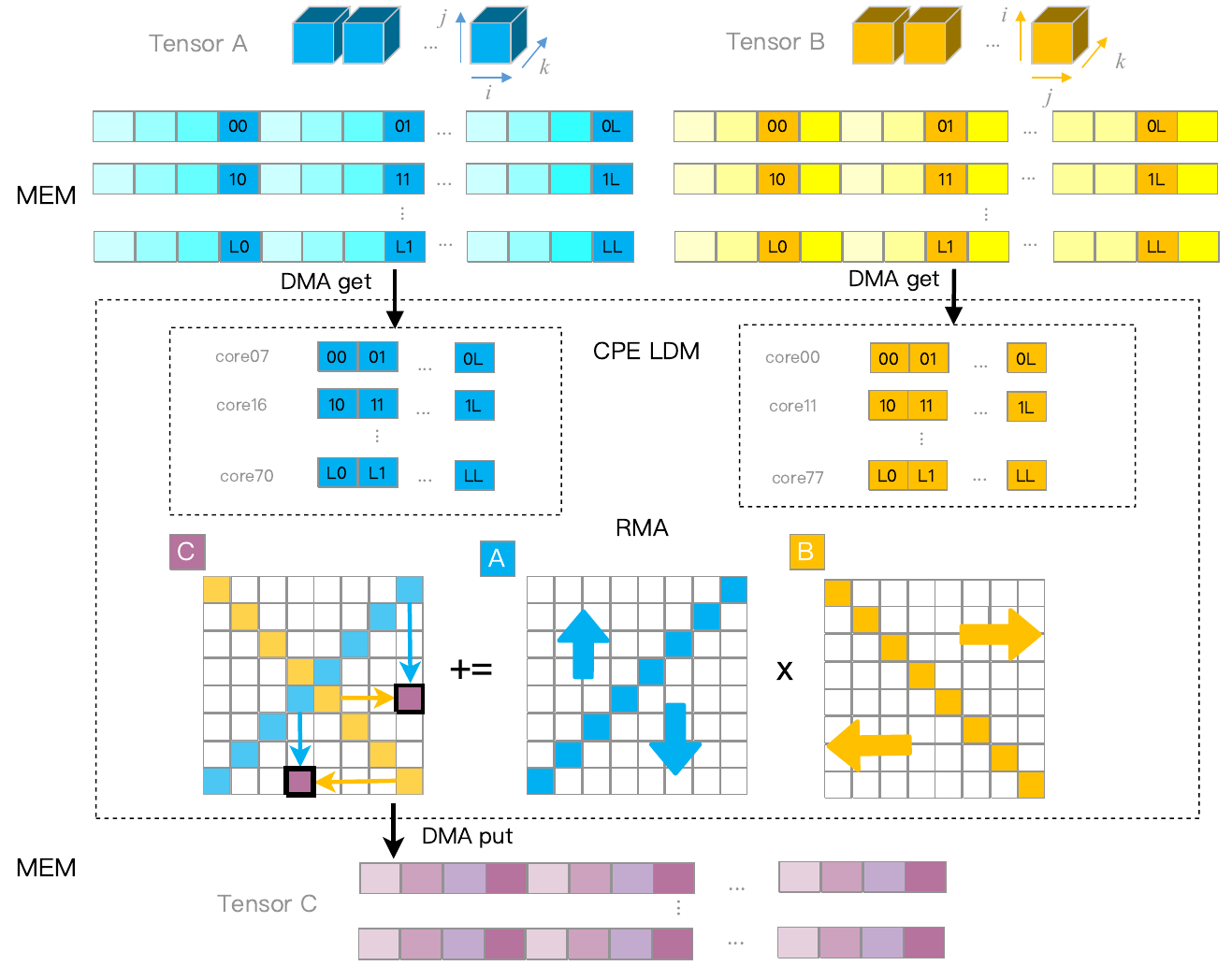}
    \caption{A fused design of tensor permutation and matrix multiplication, for tensor contractions with a high compute density.} 
    \label{fig:permutation1} 
\end{figure}

As mentioned in Section \ref{sec:sim-method} and \ref{sec:3d-path}, we now face two different kinds of tensor contractions when simulating different RQCs. 

For the PEPS-based approach, we generally have contractions between tensors with a rank around 5 or 6, and a dimension size of 32, leading to a high compute density. For such cases, we employ different CPEs to perform the fused operation in a collaborative way. As shown in Fig. \ref{fig:permutation1}, for a 2D array of CPEs, we instruct each of them to read its corresponding data block in a strided DMA pattern, so as to fetch the corresponding indices with relatively high utilization of the DDR bandwidth. Taking the tensors A and B as the example, after the strided read, the corresponding parts of A and B are loaded into the same CG of CPEs. Afterwards, we can perform a highly efficient matrix multiplication in a cooperative way within the CPE array. We use the two diagonals of the CPE array as the data broadcasters to make full use of on-chip network bandwidth. As shown in Fig. \ref{fig:permutation1}, in each step, each CPE on the diagonal performs a broadcast to forward its data to its corresponding row or column (the CPEs on the A diagonal broadcast along the columns, and the CPEs on the B diagonal broadcast along the rows),then each CPE perform the sub-multiplication of the items available and write back the corresponding sub block. After a full looping through the sub block C, the computation of matrix multiplication achieves the final result.  

In such way, we accomplish the index permutation and the multiplication in a fused manner, which would reduce a large part of the DMA load costs and most of the DMA store costs. Moreover, the carefully-scheduled diagonal broadcasting scheme achieves the best utilization of row/column buses, as well as the DMA resources, achieving a balance between the RMA and DMA efficiencies.

For the optimized contraction paths generated by the CoTenGra software in the Sycamore simulation, we usually have imbalanced contraction cases between a rank-30 tensor and a rank-4 tensor, and a much smaller dimension size of 2. In such cases, the compute density drops significantly, and the memory bandwidth becomes the main constraint. In such cases, we perform the fused operation within each CPE independently.

\begin{figure}[ht]                                    
    \centering 
    \includegraphics[width=0.75\textwidth]{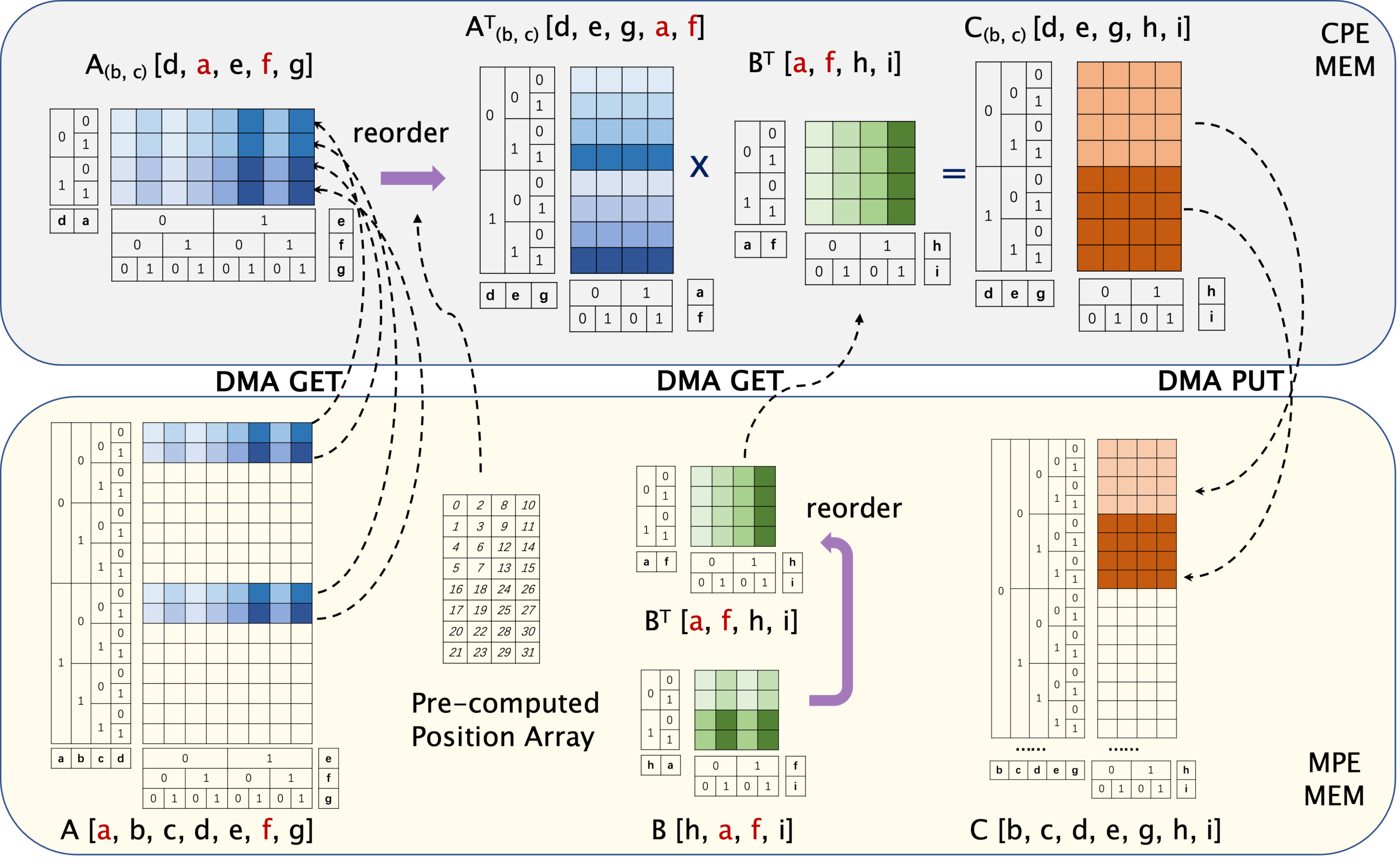}
    \caption{A fused design of tensor permutation and matrix multiplication (using the Transpose-Transpose-GEMM Transpose (TTGT) algorithm \cite{springer2018design}), so as to achieve efficient contraction for imbalanced tensor contraction cases.} 
    \label{fig:permutation2} 
\end{figure}

As shown in Fig. \ref{fig:permutation2}, we implement the tensor contration operation using the general Transpose-Transpose-GEMM Transpose (TTGT) algorithm \cite{springer2018design}, which fuses the permutation and the multiplication as one unified workflow. 

Let $A[a,b,c,d,e,f,g]$ be a tensor with $m=7$ indices and $B[h,a,f,i]$ be a tensor with $n=4$ indices, each with two dimensions. Let ${a,f}$ be the $s=2$ common indices to be contracted. Let$C[b,c,d,e,g,h,i]$ be the contraction product. 

With $B$ as the smaller tensor, we permutate it as $B^T[a,f,h,i]$ and store it in the LDM fast buffer. For the high-rank tensor $A$, the permutation from ${a,b,c,d,e,f,g}$ to ${b,c,d,a,e,f,g}$ is distributed across different CPEs. Each CPE reads a contiguous block of the last $k-s$ indices using DMA, with the size of $2^{k-s}$ ($k=5$ in this example, conditioned by the LDM size). The reading positions are determined by the new order ${b,c,a,d}$ of the remained indices ${a,b,c,d}$. The permutation of the inner indices ${d,a,e,f,g}$ to the new order ${d,e,g,a,f}$ could be performed inside LDM with a pre-computed position array to avoid repetitive memory address calculation. Then we apply a fully optimized GEMM kernel within the LDM to produce the partial result of C with indices order {d, e, g, h, i}, and transfer a contiguous data block of the size $2^{k+n-2s}$ to the main memory through DMA.

\subsection{A Mixed-Precision Computation Method via Adaptive Precision Scaling}
\label{sec:mixed}

An advantage in classical quantum simulation is that we do not have uncontrollable noises. As independent contractions to compute a single amplitude can be considered as orthogonal paths that contribute equally to the final amplitude, computing a fraction $f$ of paths is considered as equivalent to computing noisy amplitudes of fidelity $f$ \cite{markov2018quantum,villalonga2020establishing}.

Following such an idea, we propose a mixed-precision scheme that uses both single-precision and half-precision floating-point numbers to simulate RQCs. 

First, we do a pre-analysis to detect the precision sensitivity in different parts of the computation. We perform a small portion of the tensor computation to evaluate the degree of sensitivity to the switch from single to half precision. The pre-analysis results show that the parts close to the slicing positions are more sensitive to the precision reduction.

Second, we design an adaptive scaling method in order to adjust the data dynamically, especially for the precision-sensitive parts, so as to keep the error at a similar level to single-precision computation. Through the analysis of the tensor's accuracy range, a dynamic strategy for data scaling is proposed to effectively prevent data underflow, offering a better balance on precision and efficiency.

Third, we add a filter at the end of contraction. For the computation of each amplitude, taking the $10\times10\times(1+40+1)$ circuit shown in Fig. \ref{fig:tensor-slicing} as an example, we have $32^6$ paths to compute. Over all the paths, we keep the effective results without underflow exceptions, and filter out the cases with underflow or overflow cases. Due to the customized floating-point format, and the adaptive scaling method, the underflow and overflow cases are less than 2\% of the total cases, meaning that only a small fraction of the mixed-precision runs get discarded.

\begin{figure}[ht]                                    
    \centering 
    \includegraphics[width=0.75\textwidth]{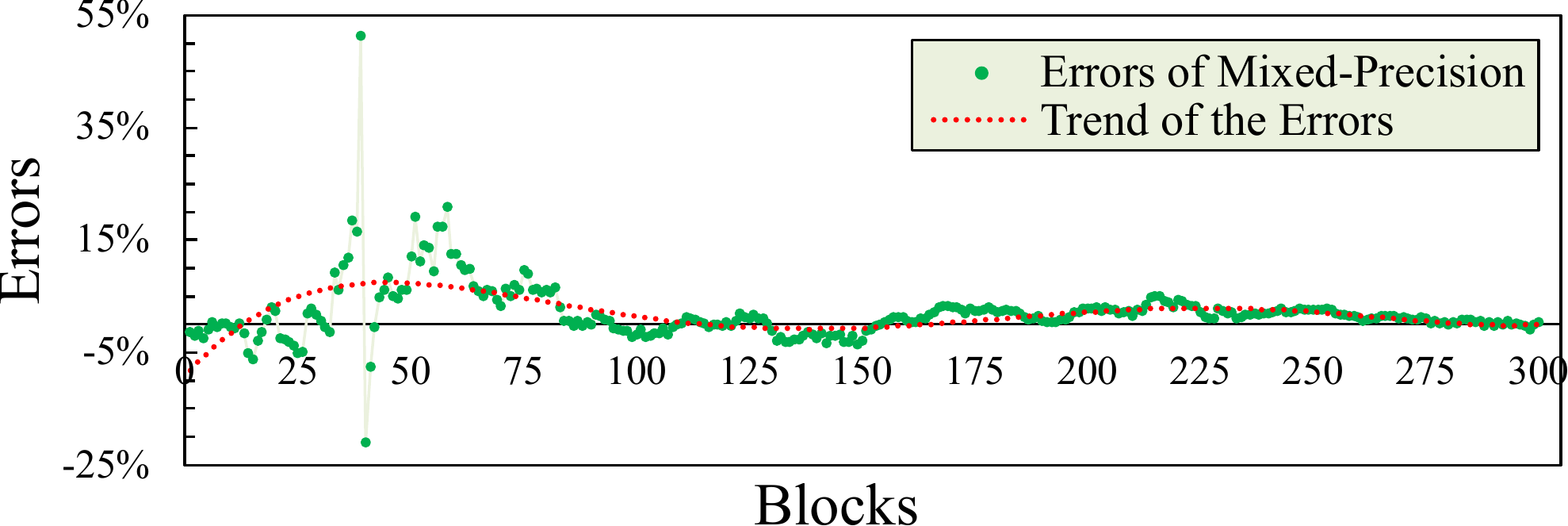}
    \caption{Errors of mixed-precision simulations, when compared with the single-precision simulations. The dotted red line shows the convergence trend of the error when computing more and more blocks (each block corresponds to 90 different contraction paths).} \label{fig:mixed} 
\end{figure}

Fig. \ref{fig:mixed} shows the errors of mixed-precision simulations of the $10\times10\times(1+40+1)$ circuit, when compared against the single-precision simulations. The results demonstrate a clear convergence of the error when we accumulate more and more paths. With around 300 blocks (27,000 paths) aggregated together, the error drops within 1\%.

For the simulation of the Sycamore, we take a different mixed-precision approach. As the memory bandwidth becomes the major bottleneck for the tensor contractions in that case, we store the variables in half-precision formats, and perform the computation in single-precision, so as to further boost the performance under the same memory bandwidth constraint.

\section{Performance Results}

\subsection{How Performance Was Measured}

The performance is measured using the average time recorded for running the same case for three times. The number of floating point operations are measured using two different methods, by counting all floating point arithmetic instructions needed for the matrix permutation and multiplication operations, and by monitoring the floating-point operation hardware counters in the processor. As the hardware counters generally provide a number that is 10$\sim$20\% larger (due to the generation of temporary floating-point operations along the way), we use the counted number as the basis for a  conservative and fair Flop measurement. 

\subsection{Result Validation with Different Precisions}

\begin{figure}[ht]                                    
    \centering 
    \includegraphics[width=0.6\textwidth]{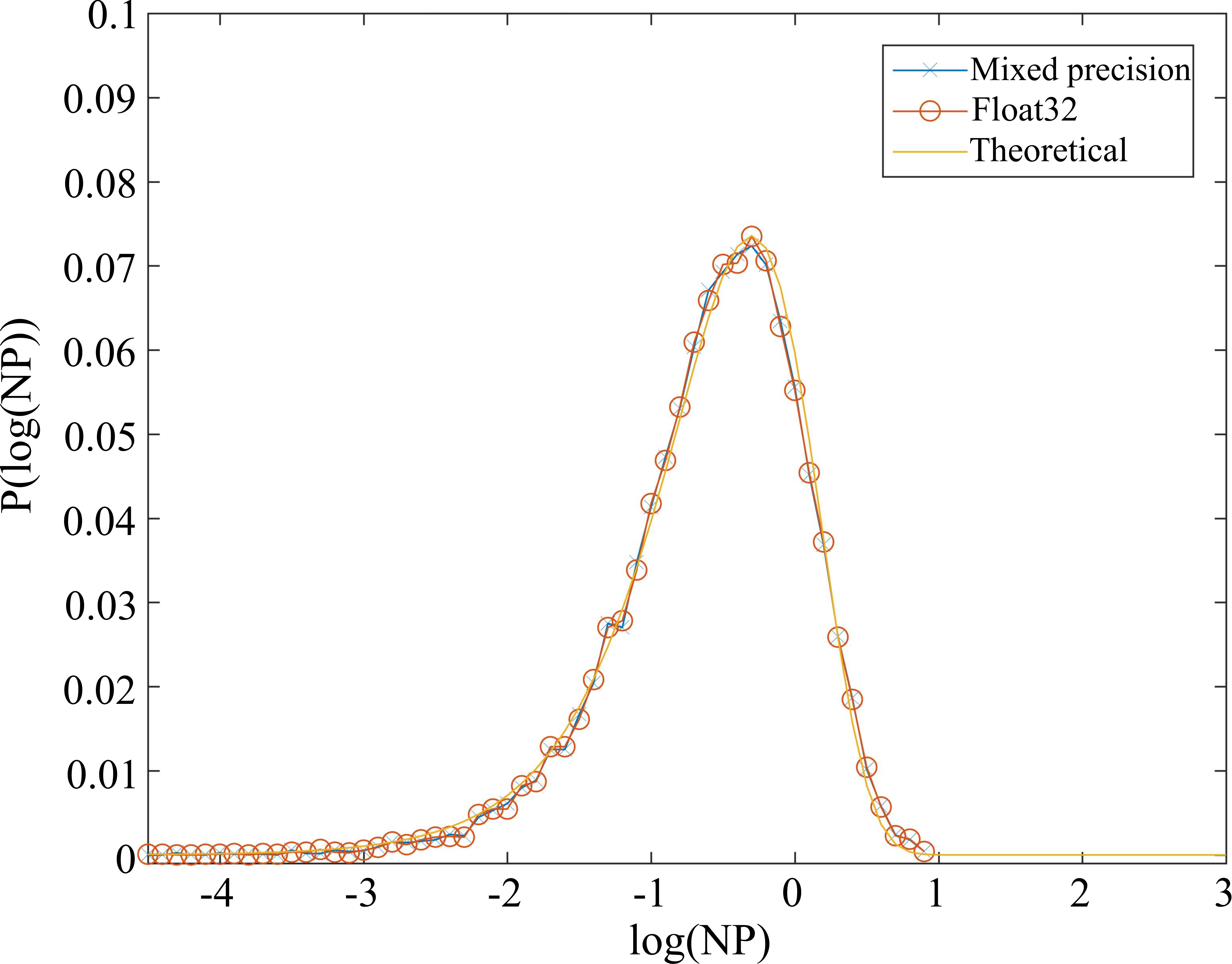}
    \vspace{-4mm}
    \caption{The probabilities for calculating 12,288 amplitudes. The red dots and blue crosses show the simulated results using single and mixed-precision floating-point numbers respectively, while the solid line shows the theoretical Porter-Thomas distribution.} \label{fig:validation} 
\end{figure}

To validate the accuracy of our proposed RQC simulator, we simulate the $10\times10\times(1+16+1)$ RQC with 12,288 amplitudes. Fig. \ref{fig:validation} plots the frequency with which each probability of configurations appear, when using single-precision and mixed-precision floating-point numbers. The red dots and blue crosses correspond to the results simulated using single and mixed-precision floating-point numbers respectively, and the line shows the theoretical Porter-Thomas distribution. The comparison demonstrates that both our single-precision and mixed-precision results fit well with the expected theoretical results. From a statistical point of view, the single-precision and mixed-precision simulations demonstrate a similar level of fidelity.

\subsection{Computing Performance and Efficiency within Each Node}
\label{sec:node-perf}

\begin{figure}[ht]                                    
    \centering 
    \includegraphics[width=0.75\textwidth]{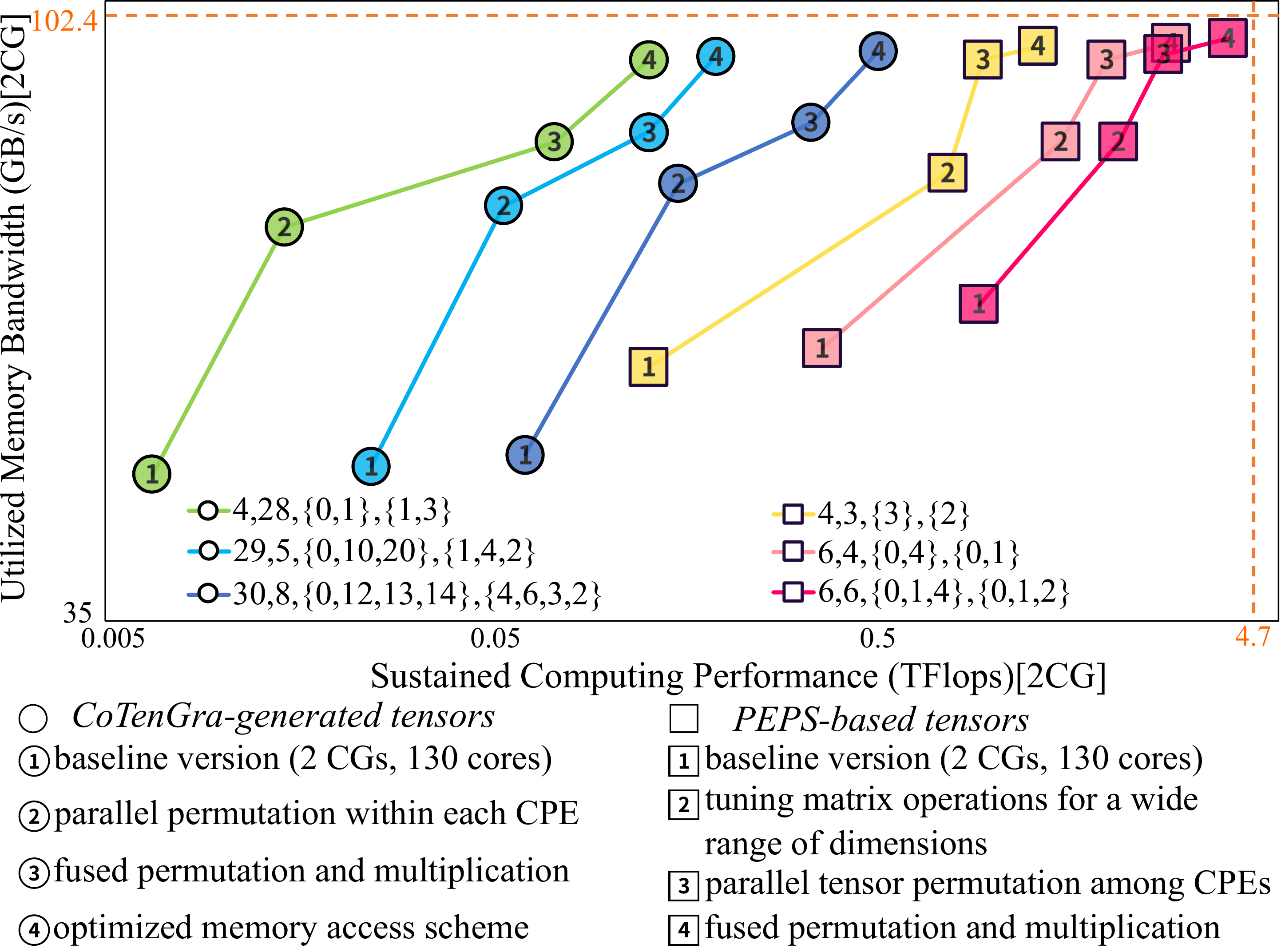}
    \caption{Computing performance of fused index permutation and multiplication operations with each MPI process (corresponding to a pair of CGs). The dashed lines indicate the peak computing performance and memory bandwidth of two CGs.} \label{fig:fuse-perf} 
\end{figure}

With the tensor contraction being the major computing task, the efficiency of the fused tensor permutation and multiplication kernels largely determines the overall performance. Fig. \ref{fig:fuse-perf} shows the computing performance and the memory bandwidth utilization of the fused index permutation and multiplication operations for a number of different tensor contraction scenarios. 

For the tensor cases frequently used in our PEPS-based approach (with most ranks around 5 and a dimension size of 32, demonstrating an apparently compute-intensive pattern), the fused permutation and multiplication kernels demonstrate an ideal compute to memory ratio, and achieve a performance that is close to the peak of 4.4 Tflops, providing a high efficiency of over 90\%. 

For the tensor cases in the optimized contraction paths provided by the CoTenGra software in the Sycamore simulation (usually between a rank-30 tensor and a rank-4 tensor, but a much smaller dimension size of 2), the compute density drops significantly, leading to a clear memory-bound pattern as shown in Fig. \ref{fig:fuse-perf}. While the sustained performance is significantly lower than the dense tensor contractions in the PEPS approach (0.2 Tflops v.s 4.4 Tflops), the optimization schemes achieve a close-to-full utilization of the available memory bandwidth. 

Among the different optimization techniques, the handling of the permutation, including both the parallelization and the fused design with multiplication, contributes most performance improvement. For the memory-bound cases generated by CoTenGra, we do further memory optimization by aggregating the load and store requests from different CPEs. 

In general, our proposed method and tuning scheme can support a wide range of tensor cases, and derive their most efficient implementation on the Sunway many-core processor, thus enabling highly efficient simulation of RQCs with different patterns and features.   

\begin{figure}[ht]                                    
    \centering 
    \includegraphics[width=0.75\textwidth]{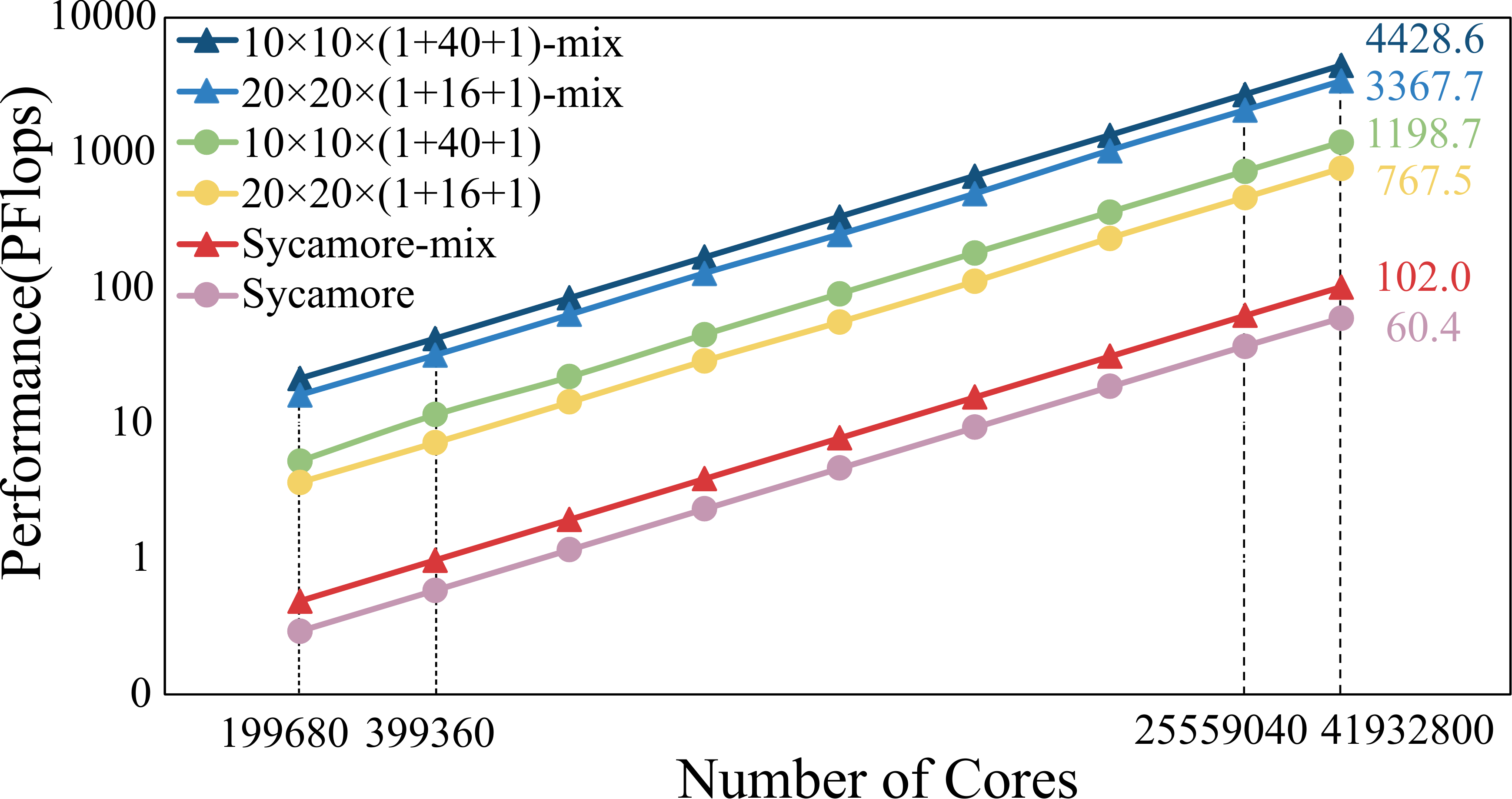} 
    \caption{The strong scaling results for a number of different complex RQCs, scaling from 199680 cores (512 computing nodes) to 41,932,800 cores (107,520 computing nodes).} \label{fig:strong-scaling}   
\end{figure}

\subsection{Scaling}

An important advantage of our method is to achieve a large number of parallel tasks to process independent sliced tensors that can be distributed to a large number of nodes. We do a global reduction at the end to collect the results and output the probability. 

Fig. \ref{fig:strong-scaling} shows the strong scaling results for simulating three different kinds of RQCs ($10\times10\times(1+40+1)$, $20\times20\times(1+16+1)$, and Sycamore), with both a single-precision and a mixed-precision configuration. All different cases demonstrate a nearly linear scaling pattern, due to the parallel-friendly feature of the slicing scheme. Among different circuits, the ones with a larger depth generally involve a higher density of tensor operations, thus providing a higher performance at different parallel scales. The highest performance comes from the simulation of the $10\times10\times(1+40+1)$ circuit, providing a single-precision performance of 1.2 Eflops, and a mixed-precision performance of 4.4 Eflops. 

Compared with the 2N by 2N lattice circuits, simulation of Sycamore demonstrates a relatively low efficiency on utilizing the Sunway machine, mainly due to the generated imbalanced tensor contraction cases discussed in Section \ref{sec:node-perf}.

\section{Implication}

This paper presents our efforts on developing an RQC simulator on the new Sunway Supercomputer. By combining both algorithmic and architecture-related optimizations, we manage to achieve an efficient mapping from very complex square quantum circuits (ranging from 10$\times$10 qubits with a depth of (1+40+1) to 20$\times$20 qubits with a depth of (1+16+1)) to the new Sunway supercomputer with around 42 million cores. 

With the slicing scheme generating enough parallelism to feed the 42 million cores, the contraction of high-rank tensors, involving index permutations and matrix multiplication, becomes the major optimization target. Our proposed strategy to fuse the permutation and multiplication into a unified workflow, manages to remove or hide most of the data movement overhead in high-rank permutations, and improves the computing efficiency by around 40\%, for both compute-intensive and memory-bound contraction cases. Moreover, the adoption of a mixed-precision can further improve the performance by over 3 times (from 1.2 Eflops to 4.4 Eflops).   

As far as we know, this is by far the highest performance achieved for simulating an RQC that is also the largest scale ever. Table \ref{tab1} compares our work with other related efforts in terms of sustained floating-poing performance and time to solution for sampling the Google Sycamore system. Compared with the previous performance of 281 Pflops on Summit for simulating a $7\times7\times(1+40+1)$ circuit, we can now accomplish the simulation of a $10\times10\times(1+40+1)$ circuit with a sustained performance of 1.2 Eflops (single-precision) or 4.4 Eflops (mixed-precision). For the case of simulating Sycamore, even though the generated tensors demonstrate an imbalanced contraction scenario and bring tough challenge for the memory system, our simulator, with memory optimization for a wide range of tensor cases, can still provide an efficiency of 4\%. Compared with all the extreme-scale simulation or data analysis work on leading-edge systems in recent years \cite{jia2020pushing,joubert2018attacking,kurth2018exascale}, our work provides by far the highest sustained floating-point performance and efficiency.

For the specific case of simulating the Google Sycamore system, with all the algorithmic innovations and architecture-related optimizations introduced above, we manage to accomplish the sampling task within 304 seconds, demonstrating a real-time simulation capability for the most advanced quantum circuits of current days. Moreover, the current performance is largely constrained by the relatively low efficiency for contracting the tensors generated by the CoTenGra software. We believe a customization of the code to generate more balanced tensors for the Sunway system could further improve the speed by another factor of 4 to 5 times.

The current path we use for tensor contraction in the Sycamore simulation, and the simulation results of two million different amplitudes can be accessed at \url{https://github.com/Daisyforest/SWQsim}.

\begin{table}[ht]
\centering
\footnotesize
\caption{Comparing our work with related efforts on floating-point performance and quantum circuit sampling speed.}
\begin{tabular}{lll}
\midrule
 \multicolumn{3}{c}{Computational performance and efficiency} \\ \hline
 &  FLP32 & FLP16 (mixed)  \\\hline
{\bf our simulation} of & \multirow{2}{*}{$\frac{1.2 Eflops}{1.5 Eflops}\approx80.0\%$} & \multirow{2}{*}{$\frac{4.4 Eflops}{5.9 Eflops}\approx74.6\%$} \\
10x10x(1+40+1) & & \\\hline 
{\bf our simulation} of & \multirow{2}{*}{$\frac{60.4 Pflops}{1.5 Eflops}\approx4.0\%$} & \multirow{2}{*}{$\frac{102 Pflops}{5.9 Eflops}\approx1.7\%$} \\
Sycamore  & & \\\hline
qFlex on Summit & \multirow{2}{*}{$\frac{281 Pflops}{415 Pflops}\approx67.7\%$} & \multirow{2}{*}{n/a} \\
7x7x(1+40+1) & & \\\hline
MD on Summit with  & \multirow{2}{*}{$\frac{162 Pflops}{415 Pflops}\approx39.0\%$} & \multirow{2}{*}{$\frac{275 Pflops}{3.3 Eflops}\approx8.3\%$} \\
machine learning \cite{jia2020pushing} & & \\\hline
genetic data & \multirow{2}{*}{n/a} & \multirow{2}{*}{$\frac{2.3 Eflops}{3.3 Eflops}\approx69.7\%$} \\
analysis on Summit \cite{joubert2018attacking} & & \\\hline
deep learning for & \multirow{2}{*}{n/a} & \multirow{2}{*}{$\frac{1.13 Eflops}{3.3 Eflops}\approx34.2\%$} \\
climate on Summit \cite{kurth2018exascale} & & \\\hline\hline
\multicolumn{3}{c}{Time needed to sample Sycamore} \\ \hline
{\bf our simulation} & 304 seconds & \\\hline
physical Sycamore \cite{google-nature-2019} & 200 seconds & \\\hline
Summit\cite{google-nature-2019} & 10,000 years & \\\hline
Summit\cite{pednault2019leveraging} & 2.55 days (estimated) & \\\hline
Ali\_Cloud\cite{huang2020classical} & 19.3 days (estimated)  \\ \hline
60\_GPUs(Pan)\cite{pan2021simulating} & 5 days \\

\midrule
\specialrule{0em}{1.5pt}{1.5pt}
\midrule
      
\end{tabular}
\label{tab1}
\end{table}

An important lesson we learnt from this work is the huge potential that emerging new disciplines could bring to the development of algorithms and optimizations in the supercomputing community. With a strong motivation to demystify and to narrow down the ``Quantum Supremacy'' of emerging quantum computing systems, the methods and optimization techniques related to classical simulation of RQC develop fast in the last few years. Combining all algorithmic improvements in previous efforts, as well as our innovations introduced in this paper (including strategies to identify better slicing schemes and contraction paths, and a fused workflow of tensor permutation and multiplication), we are now capable of simulating the sampling process of quantum systems at the time scale of seconds, instead of years.  

We believe such a simulator that can provide results in real-time would in return bring concrete support for the development of future quantum systems and potentially accelerate the process to design and implement more complex and more practical quantum computing systems.

\section{Acknowledgement}

We would like to thank Dapeng Yu, Heng Fan, Guoping Guo, Yongjian Han, and Xiaobo Zhu for advices and discussions. 

This work is partially supported by National Key R\&D Program of China (2017YFA0604500), and National Natural Science Foundation of China (U1839206). The corresponding authors are Xin Liu (lucyliu\_zj@163.com), Haohuan Fu (haohuan@tsinghua.edu.cn), Chu Guo (guochu604b@gmail.com), and Dexun Chen (adch@263.net).

\bibliographystyle{plain}
\bibliography{ref}

\appendix

\section{Technical Details on Comparing Our Simulator Against the Google Sycamore Quantum Processor}

Due to the page limit of our previous submission to the SC'21 conference proceeding, certain parts of the work are not fully covered. In this part, we provide some more technical details regarding the comparison of our RQC simulator on a new generation Sunway supercomputer and the Google Sycamore quantum computer \cite{google-nature-2019}. 

Quantum processors are inherently suitable for sampling tasks. By law of quantum physics, performing a measurement on a quantum processor, one would obtain a sample (bitstring) with certain probability determined by the underlying quantum circuit. The most difficult sampling task performed by Sycamore is to sample $1$ million bitstrings from a random quantum circuit of $20$ cycles, with the XEB (cross entrophy benchmarking) fidelity estimated to be $0.2\%$~\cite{google-nature-2019}. The low fidelity mainly originates from the noise of current quantum processor. A $0.2\%$ fidelity out of $1$ million samples ensures that we have $2,000$ valid samples out of the one million, although we can not tell the validity of each single sample. To estimate the computational cost of the same task on classical supercomputers, Ref.~\cite{markov2018quantum} suggests a scaling of the computational cost by a factor of the XEB fidelity, namely the classical computational cost of generating one million samples with $0.2\%$ XEB fidelity would be equivalent to that of generating $2,000$ perfect ones.


In comparison with the quantum processors which can directly generate samples, a more natural task for classical computer to do is to compute the amplitude (thus the probability) for a given bitstring (or the amplitudes for a bunch of bitstrings). To mimic the sampling task, classical computers have to perform additional sampling process among a set of bitstrings (frugal sampling for example) based on their amplitudes. Nevertheless, to produce a bunch of bitstrings whose XEB fidelity matches the quantum processor, Pan and Zhang \cite{pan2021simulating} proposes a big-head algorithm such that one could use the classical computer to generate a correlated but large bunch of amplitudes (one could perform sampling among these bitstrings afterwards), by fixing a subset of the qubits and then exhausting the rest. For the $53$-qubit Sycamore processor, they fix $32$ qubits to be $0$ (the choices of the specific values does not affect the classical simulation complexity) and then exhaust the rest $2^{21}$ qubits. The advantage is that one could obtain $2^{21}$ (about $2$ millions) exact amplitudes with almost the same classical computational complexity as that of computing a single amplitude since a major portion of the tensor contractions can be reused for those amplitudes. The XEB value of the corresponding bitstrings obtained in this way could be higher than that of the quantum processor. 

In our work, we follow the idea of Ref.~\cite{pan2021simulating} to compute a single large bunch of $2^{21}$ correlated amplitudes using the new generation Sunway supercomputer (but we did not use the big-head algorithm), with the total runtime to be $304$ seconds using mixed-precision arithmetic. The main goal is to demonstrate a parallelization scale of more than $40,000,000$ cores on the Sunway supercomputer as well as the performance gain compared to the previous approach. Identifying a most fair comparison between the sampling task of a quantum computer, such as Sycamore, and the simulation task on a traditional computer, is not our main focus of this work. Nevertheless we can give some comments here based on a few existing literatures on this topic. 

First and for a naive comparison, one could simply scale the runtime by a factor of $2,000$. Such a scaling multiplies 304 seconds into 7 days, which is a longer but still practical time frame for traditional computers. Meanwhile, $2,000$ perfect samples (with their exact probability information) are of course much more informative than one million samples with an estimated $0.2\%$ fidelity. Second, in a following work~\cite{PanZhang2021b}, it is further claimed that it is possible to generalize the big-head approach such that one could generate one million uncorrelated bitstrings, but still with a classical simulation complexity which is essentially the same as the case of correlated ones~\cite{pan2021simulating}. Last, even for computing a bunch of totally uncorrelated bitstrings, it is demonstrated that one could still reuse a major portion of the intermediate results during contracting the tensor networks corresponding to those bitstrings similar to the case of correlated ones, with speedups ranging from $20{\rm x}$ to $10,000{\rm x}$ (for random quantum circuits with cycles from $12$ to $16$) compared to the naive approach~\cite{KalachevYung2021}.

In our numerical experiment, we have randomly fixed $32$ qubits (see the first column TABLE.~\ref{tab:tab1} for those fixed qubits) and exhaust the rest ones. The XEB value corresponds to those bitstring is found to be $0.741$. We list $5$ amplitudes together with the corresponding bitstrings in TABLE.~\ref{tab:tab1} for references. 


\begin{table}[!htb]
\centering
\caption{The amplitudes corresponding to the $5$ selected bitstrings computed within a bunch of $2^{21}$ bitstrings. The qubits marked red in the bitstrings are fixed.
}
\label{tab:tab1}
\begin{tabular}{|c|c|c|c|}
\hline
bitstring &  amplitudes   \\
\hline
0000000000\mred{01}000000\mred{1110}0000\mred{1}0\mred{0101001010101010010100110} &   $2.85\times 10^{-9} - 7.23\times 10^{-9}\im$      \\
0000000000\mred{01}000000\mred{1110}0000\mred{1}1\mred{0101001010101010010100110} &   $1.60\times 10^{-8} - 6.49\times 10^{-9}\im $    \\
0000000000\mred{01}000000\mred{1110}0001\mred{1}0\mred{0101001010101010010100110} &   $-4.02\times 10^{-9} + 1.17 \times 10^{-9}\im $    \\
0000000000\mred{01}000000\mred{1110}0001\mred{1}1\mred{0101001010101010010100110} &   $-5.24\times 10^{-9} - 4.81 \times 10^{-9}\im $   \\
0000000000\mred{01}000000\mred{1110}0010\mred{1}0\mred{0101001010101010010100110} &   $-4.46\times 10^{-9} - 3.34 \times 10^{-9}\im $  \\
\hline
\end{tabular}
\end{table}

\end{document}